\documentclass[smallextended]{svjour3}       
\smartqed  
\usepackage{graphicx}
\usepackage{adjustbox} 
\usepackage{amssymb} 
\usepackage{amsmath} 
\usepackage{floatrow} 
\usepackage{nameref} 
\usepackage{floatrow}
\journalname{Journal of Low Temperature Physics}

\begin{document}

\title{Very low resistance Al/Cu joints for use at cryogenic temperatures}

\author{S\'ebastien Triqueneaux        
\and James Butterworth
\and Johannes Goupy
\and Cl\'ement Ribas 
\and David Schmoranzer 
\and Eddy Collin
\and Andrew Fefferman 
}

\institute{S. Triqueneaux \and J. Goupy \and C. Ribas \and E. Collin \and A. Fefferman \at
 Univ. Grenoble Alpes, CNRS, Grenoble INP, Institut Néel, 38000 Grenoble, France \\
\email{sebastien.triqueneaux@neel.cnrs.fr}           
\and J. Butterworth
\at Air Liquide Advanced Technologies 2, rue de Cl\'emenci\`ere, BP 15, 38360 Sassenage, France
\and D. Schmoranzer
\at Charles University, Ke Karlovu 3, 121 16, Prague, Czech Republic
}

\date{Received: date / Accepted: date}

\maketitle

\begin{abstract}
We present two different techniques for achieving low resistance ($<$20~n$\rm \Omega$) contacts between copper and aluminium at cryogenic temperatures. The best method is based on gold plating of the surfaces in an e-beam evaporator immediately after Ar plasma etching in the same apparatus, yielding resistances as low as 3~n$\rm \Omega$ that are stable over time. The second approach involves inserting indium in the Al/Cu joint. For both methods, we believe key elements are surface polishing, total removal of the aluminum oxide surface layer, and temporary application of large (typ. 11 kN) compression forces. Such contacts are not demountable. We believe the values for gold plated contacts are the lowest ever reported for a Cu/Al joint of a few $\rm cm^{2}$. This technology could simplify the construction of thermal links
for advanced cryogenics applications, in particular that of extremely low resistance heat switches for nuclear demagnetization refrigerators.

\end{abstract}

\section{Introduction}
\label{intro}
Dry dilution refrigerators have widely spread in laboratories these last years, opening the mK temperature range to a large community. Yet, some research fields still require lower temperatures. As an illustration, a European project has recently been granted~\cite{Pickett2018} aiming at providing access to platforms in the micro-kelvin range. Nuclear demagnetization refrigerators (NDRs) pre-cooled by dry dilution fridges seem a straightforward extension of the temperature range below 1 mK. Successful developments~\cite{Todoshchenko2014,Batey2013} either with Cu or with PrNi$_5$ as nuclear coolant have validated this concept.

NDRs are not intrinsically continuous coolers as they need to be recycled. In wet dilution refrigerators, i.e. those using liquid helium as a 4.2 K precooling source, nuclear stages can remain at base temperature typically for a couple of days or, in some cases, for up to a month (see~\cite{Pobell2007} and references therein). In these examples, the heat leak is typically well below 1~nW. In dry fridges, the induced vibrations associated with pulse-tube coolers~\cite{Riabzev2009,Schmoranzer2019} can lead to a significant heat input and values of the order of 5~nW have been reported~\cite{Todoshchenko2014,Batey2013}. For a given setup, this will shorten the experimental time at base temperature. For this main reason, but also since operation in the sub-mK regime for months would be of interest for some experiments, there is a need for the development of a continuous nuclear demagnetization refrigerator (CNDR) as already proposed by Toda et al.~\cite{Toda2018}.

\begin{figure}[h!]
\includegraphics[width=0.88\linewidth]{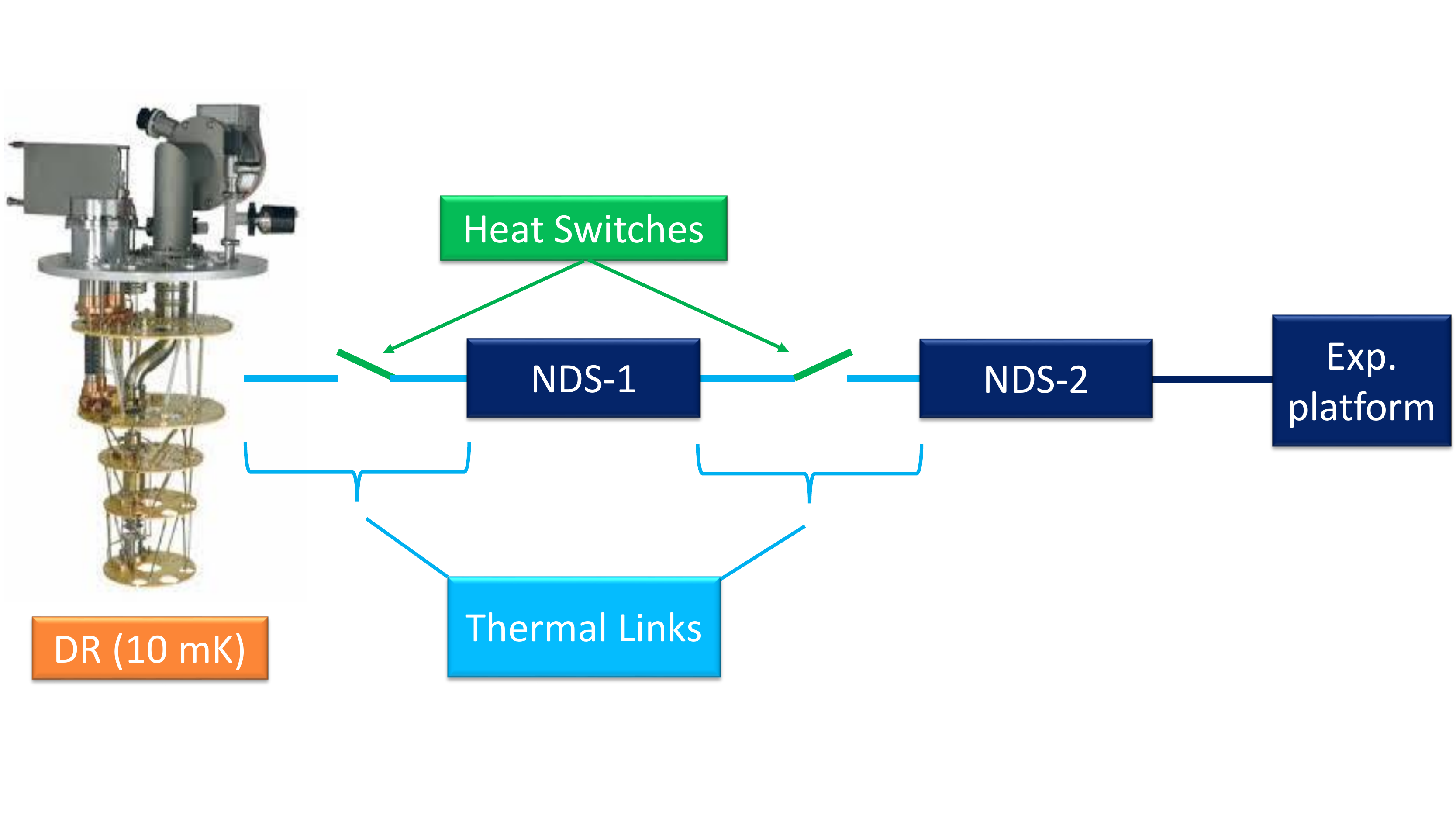}
\caption{Overview of the Continuous Nuclear Demagnetisation Refrigerator (CNDR). DR stands for Dilution Refrigerator and NDS stands for Nuclear Demagnetisation Stage. Note that, for clarity, the coils around the NDSs and the superconducting heat switches are not represented.}
\label{fig:Overview}       
\end{figure}

In a previous paper~\cite{Schmoranzer2019-2}, we have presented a CNDR concept based on two PrNi$_5$ nuclear demagnetization stages (NDSs) mounted in series (see Fig.~\ref{fig:Overview}) and separated by a heat switch. This architecture, aimed at providing continuous operation below 1~mK, is similar to the one used for electronic adiabatic demagnetization refrigerators for space applications~\cite{Shirron2004}. According to our simulations~\cite{Schmoranzer2019-2}, the thermal link between the 2 stages is the most critical point: its thermal resistance must be minimized. This is detailed in the following section.

In a more recent paper~\cite{Schmoranzer2020}, we have compared parallel and series configurations. Although the parallel configuration may provide better performance and relax the constraint on the thermal link resistance, it requires more space and is more complex since it requires 4 heat switches instead of 2. Whichever the selected configuration, the thermal link resistance should be minimized.

In the present work, we report stable Al/Cu contact resistances as small as 3~n$\rm \Omega$. This achievement may allow construction of superconducting heat switches with a normal state resistance $\leq$15~n$\rm \Omega$, which is the lowest value reported so far~\cite{Mueller1978}. Furthermore, the construction of our gold-plated Al/Cu joints rely on removal of the aluminum oxide layer by plasma etching and subsequent gold deposition without breaking vacuum. Our process avoids the complicated and potentially dangerous cyanide-based chemistry employed in~\cite{Mueller1978}.

\section{Thermal link issue}
\label{sec:1}

\begin{figure}[h!]
\floatbox[{\capbeside\thisfloatsetup{capbesideposition={left,center},capbesidewidth=4cm}}]{figure}[\FBwidth]
{\caption{Final temperature (x) for different heat leaks (y) and different values of the NDS1-NDS2 thermal link electrical resistance (heat switch in its normal state) taken from \cite{Schmoranzer2019-2}. Dashed lines identify the selected parameters.}
\label{fig:TvsR}}
{\includegraphics[width=6cm]{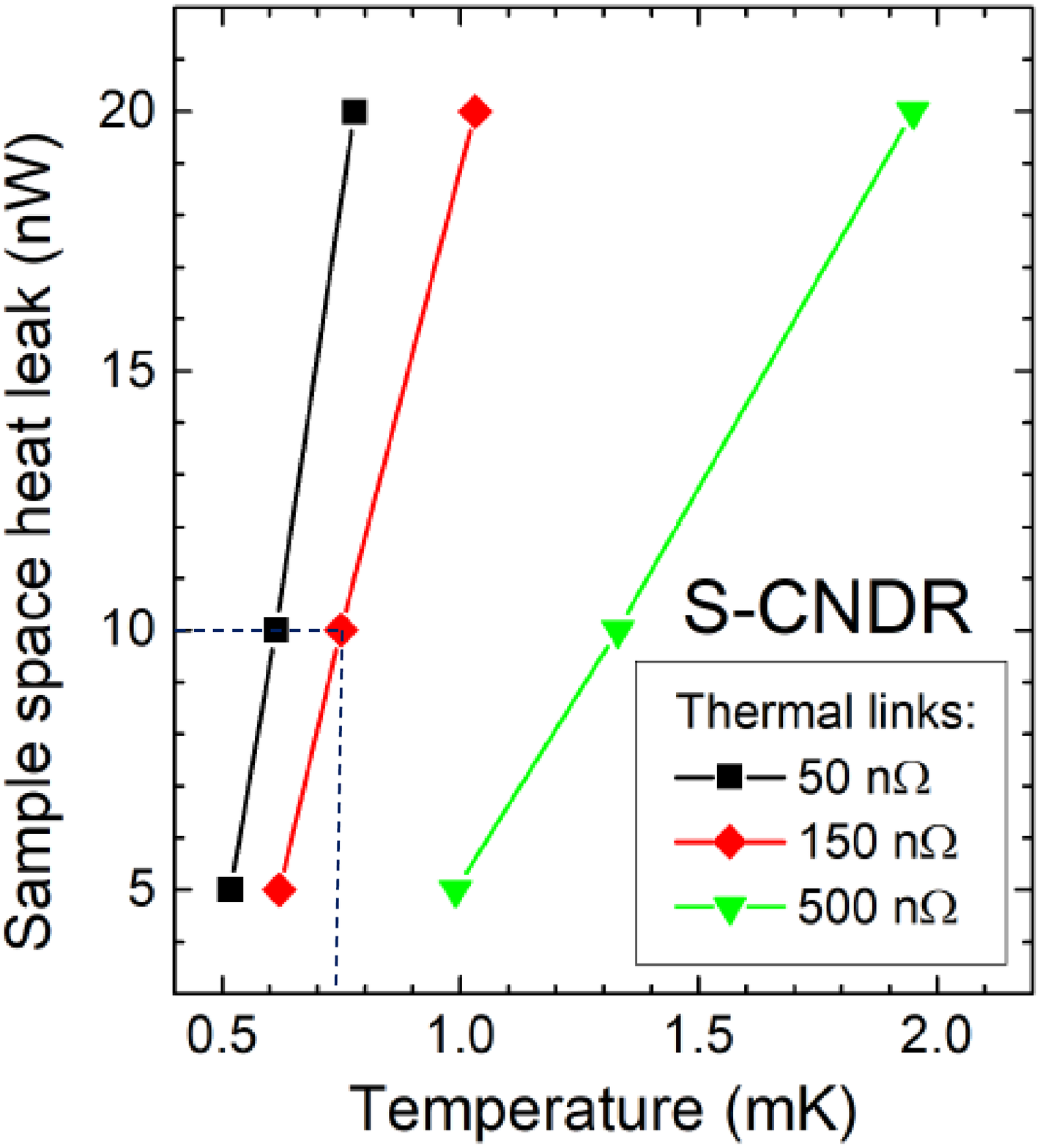}}
\end{figure}

Fig.~\ref{fig:TvsR} highlights the criticality of the thermal link between NDS1 and NDS2. According to the model developed in~\cite{Schmoranzer2019-2,Schmoranzer2020} and assuming a minimum field-independent heat leak of 5~nW, it is not possible to maintain a base temperature below 1~mK for a thermal link resistance of 500~n$\rm \Omega$. With 150~n$\rm \Omega$, the model suggests that the base temperature could be as low as 750~µK for 10~nW losses. It is expected that this provides sufficient margin to allow for uncertainties inherent to the model. As a consequence, this value of 150~n$\rm \Omega$ can be considered an appropriate objective value.

In the following, we have used the Wiedemann-Franz law to convert thermal resistances into electrical resistances. As an example, an electrical resistance of 150~n$\rm \Omega$ corresponds to a thermal resistance of $\approx$6000~K/W at 1~mK which would lead to an acceptable $\rm \Delta$T of 60~µK between the 2 NDS for 10~nW applied power. Although this law can overestimate the thermal conductivity~\cite{Shirron2004,Gloos1990} for some materials or for contact resistances, we assume it is satisfactory for comparison purposes.

In order to evaluate whether this requirement was achievable, we have reviewed some of the articles dealing with superconducting heat switches for NDRs. Results are summarized in Table~1. The equivalent electrical resistances vary over almost 4 orders of magnitude, but at least 6 systems exhibit values close to or below 100~n$\rm \Omega$ and the lowest reported value is only 15~n$\rm \Omega$. This suggests that our goal value of 150~n$\rm \Omega$ for the entire thermal link between the 2 NDSs might be achievable.

\begin{table}[p]
\label{tab:HeatSwitches}
\centering
\begin{adjustbox}{rotate=90,max width=\textwidth}
\caption{A review of heat switches thermal performance (sorted by publication date).}
		\begin{tabular}{|c|c|p{7cm}|c|p{2.8cm}|c|p{4cm}|}
		\hline
		\textbf{Ref.} & \textbf{Mat.} & \textbf{Technology} & \textbf{Variable} & \textbf{Published Value} & \textbf{Equiv. R} & \textbf{Comments} \\
		\hline
		\cite{Krusius1978} & Zn & Zinc foils soldered to copper & Thermal R. & $120/T~{\rm K^{2}W^{-1}}$ & $\approx$2900~n$\rm \Omega$ &  \\
		\hline
                     \cite{Mueller1978} & Al & Al foils pressed against copper foils + gold plating (zincate solution for Al) & Conductance  & At 30~mK: 10~$\rm \mu$W with a $\rm \Delta$T of 0.2~mK & 15~n$\rm \Omega$ & For the overal heat switch \\
		\hline
		\cite{Lawson1982} & Al & Bulk Al with silver wires - fusion under pressure & Electrical R. & $0.5~\rm \mu$$\rm \Omega$ per contact & $\approx$250~n$\rm \Omega$ & For 4 contacts in parallel and 2 in series and negligible contribution of Al or Ag wires \\
		\hline
		\cite{Schuberth1984} & Sn & Sn Switch & Conductance & $\rm 5 \times 10^{-3} T~ WK^{-2}$ & 1200~n$\rm \Omega$ & \\
		\hline
		\cite{Schuberth1984} & In & In Test switch & Conductance & 0.4 T~WK$^{-2}$ & 61~n$\rm \Omega$ &   \\
		\hline
		\cite{Gloos1988} & Al & Overall link between DR and 1$^{st}$ NDS & Conductance & 0.24 T~WK$^{-1}$ & 102~n$\rm \Omega$  &   \\
		\hline
		\cite{Gloos1988} & Zn & Five Zn foils (0.1~mm thick) pressed into Cu slits & Electrical R. & 0.4~$\rm \mu$$\rm \Omega$ & 400~n$\rm \Omega$  & Measured at 4.2~K  \\
		\hline
		\cite{Gloos1988} & Zn &Same as above & Conductance & 0.04 T~WK$^{-1}$ & 610~n$\rm \Omega$  & Between 12 and 35~mK \\
		\hline
		\cite{Gloos1988} & Al & Five Al foils (0.1~mm thick) pressed into Cu slits + tungsten washers & Electrical R. & 0.1~$\rm \mu$$\rm \Omega$  & 100~n$\rm \Omega$  &  \\
		\hline
		\cite{Gloos1988} & Al & Same as above & Conductance & 0.2 T~WK$^{-1}$ & 122~n$\rm \Omega$  & Between 1.1 and 3.8~mK  \\
		\hline
		\cite{Gloos1988} & Pb & $10~mm \times 20~mm \times 1~mm$ bolted to Cu and Ag. Overall thermal link characterised & Electrical R. & 0.8~$\rm \mu$$\rm \Omega$  & 800~n$\rm \Omega$  &  \\
		\hline
		\cite{Gloos1988} & Pb & Same as above & Conductance & 0.04 T~WK$^{-1}$ & 610~n$\rm \Omega$  & Between 18 and 35~mK  \\
		\hline
		\cite{Gloos1988} & Pb & $10~mm \times 20~mm \times 1~mm$ bolted to Cu and Ag. Shorter Ag foils. Overall thermal link characterised & Electrical R. & 0.2~$\rm \mu$$\rm \Omega$  & 200~n$\rm \Omega$   & \\
		\hline
		\cite{Gloos1988} & Pb & Same as above. Degradation with time & Conductance & 0.03 T~WK$^{-1}$  & 813~n$\rm \Omega$  & Between 4 and 10~mK. Degradation after a few cool down  \\
		\hline
		\cite{Bunkov1989} & Al & 3 Al foils (0.4~mm thick) diffusion welded to 4 Cu foils (0.2~mm thick) & Electrical R. & 400~n$\rm \Omega$ &400 ~n$\rm \Omega$  &  \\
		\hline
		\cite{Ho2000} & In & Rolled In wire soldered at each end & Conductance & $2.7 \times 10^{-4}$ T~WK$^{-1}$ &  $\approx$90000~n$\rm \Omega$ &  \\
		\hline
		\cite{Yao2000} & Al & Eight 0.25~mm thick aluminum foils diffusion welded into Cu blocks & Electrical R. & 100~n$\rm \Omega$ & 100~n$\rm \Omega$  &  \\
		\hline
		\cite{Tajima2003} & Zn & 0.25~mm Zn foil diffusion welded to 0.5~mm Cu foils + pressed contacts to Cu blocks & Electrical R. & 50~n$\rm \Omega$ & 50~n$\rm \Omega$  &  \\
		\hline
		\cite{Tajima2003} & Zn & Complete heat switch & Electrical R. & 100~n$\rm \Omega$ & 100~n$\rm \Omega$  &  \\
		\hline
		\cite{Todoshchenko2014} & Al & Seven bended aluminium foils ($50~mm \times 10~mm \times 0.5~mm$ each) diffusion welded to two copper rods & Electrical R. & 2.4~$\rm \mu$$\rm \Omega$ & 2400~n$\rm \Omega$ &  \\
		\hline
		\end{tabular}
\end{adjustbox}

\end{table}

\section{Thermal contact issue}
\label{sec:2}
\begin{figure}[h!]
\floatbox[{\capbeside\thisfloatsetup{capbesideposition={left,center},capbesidewidth=2cm}}]{figure}[\FBwidth]
{\caption{Breakdown of the thermal resistances between the 2 NDSs.}\label{fig:ThermalRs}}
{\includegraphics[width=9.2cm]{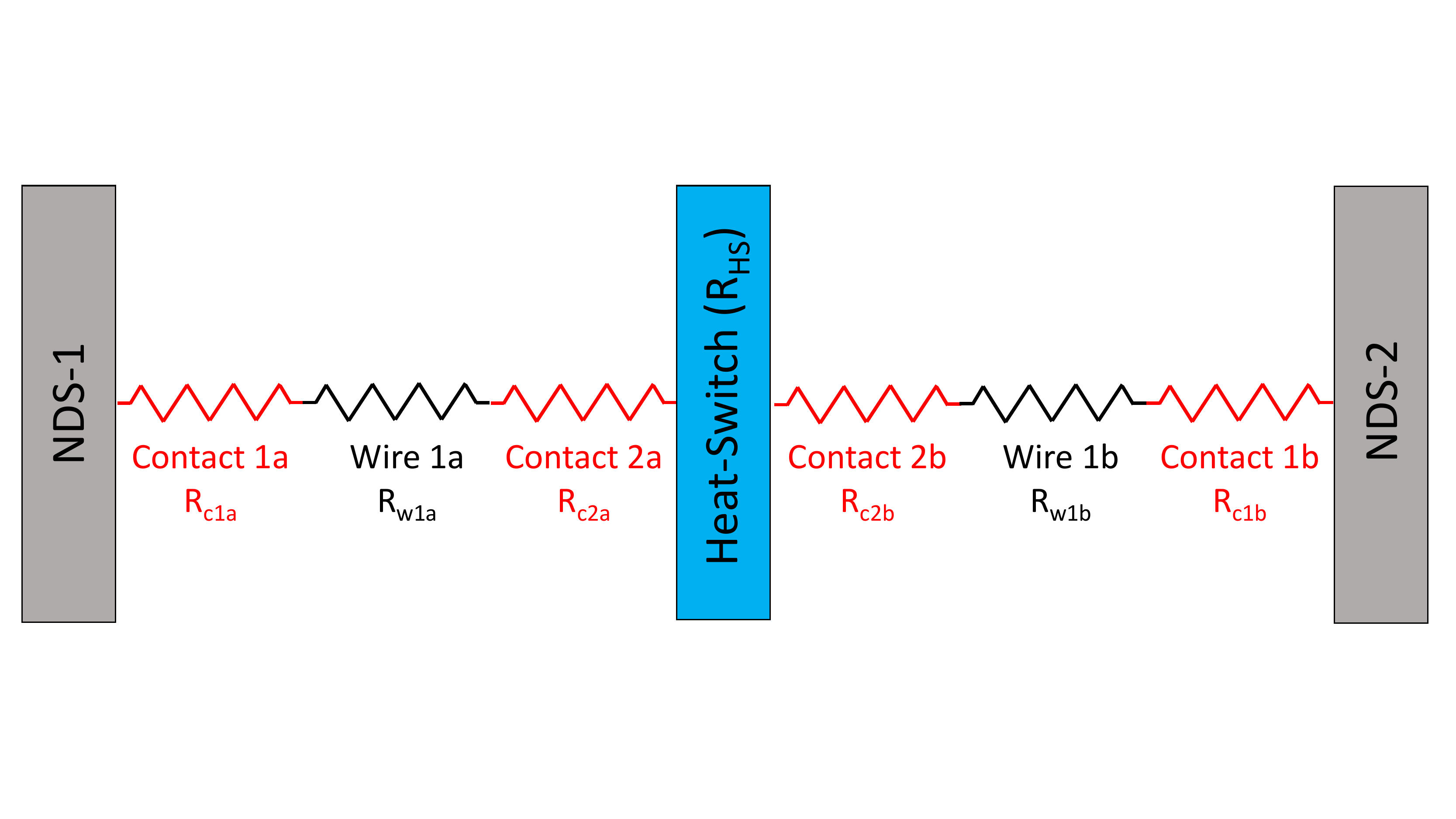}}
\end{figure}

Fig.~\ref{fig:ThermalRs} provides a breakdown of the contributors to the thermal link resistance between the 2 NDSs.

For the determination of the electrical resistances, we have used the following assumptions:

\begin{itemize}
\item Wires 1$_a$ (wires 1$_b$ are considered identical): 6 copper wires of 6N purity \cite{Alfa44551}, 1.5~mm diameter and 5~cm long. RRRs above 5 000 have been achieved several times on wires from the same batch after heat treatment. This results in 16~n$\rm \Omega$ per set of wires, so 32~n$\rm \Omega$ total contribution for R$_{w1a}$ and R$_{w1b}$.
\item The Heat-Switch will be made of Aluminium. Although its geometry is not fixed yet, it could be a C-shape block as already used at Lancaster (see \cite{Lawson1982} for example) with a section of 15~mm $\times$ 5~mm and a thermal path length of approximately 35~mm. We have measured a RRR of $\approx$5000 with 6N purity \cite{Alfa44336} as received. The calculated contribution is $\approx$2~n$\rm \Omega$.
\end{itemize}

As a result, and after applying a factor of~2 as a safety margin on the contributions of the Cu wires and Al heat switch, we end up with a budget of $\approx$20~n$\rm \Omega$ for each of the 4 contact resistances. Once again the question arises whether such a value is achievable. We have first focused our efforts on the Al/Cu contact resistances (R$_{c2a}$ and R$_{c2b}$ in Fig.~\ref{fig:ThermalRs})  leaving the PrNi$_5$-Cu contact for a future step.

\begin{figure}[h!]
\centering
\includegraphics[width=0.99\linewidth]{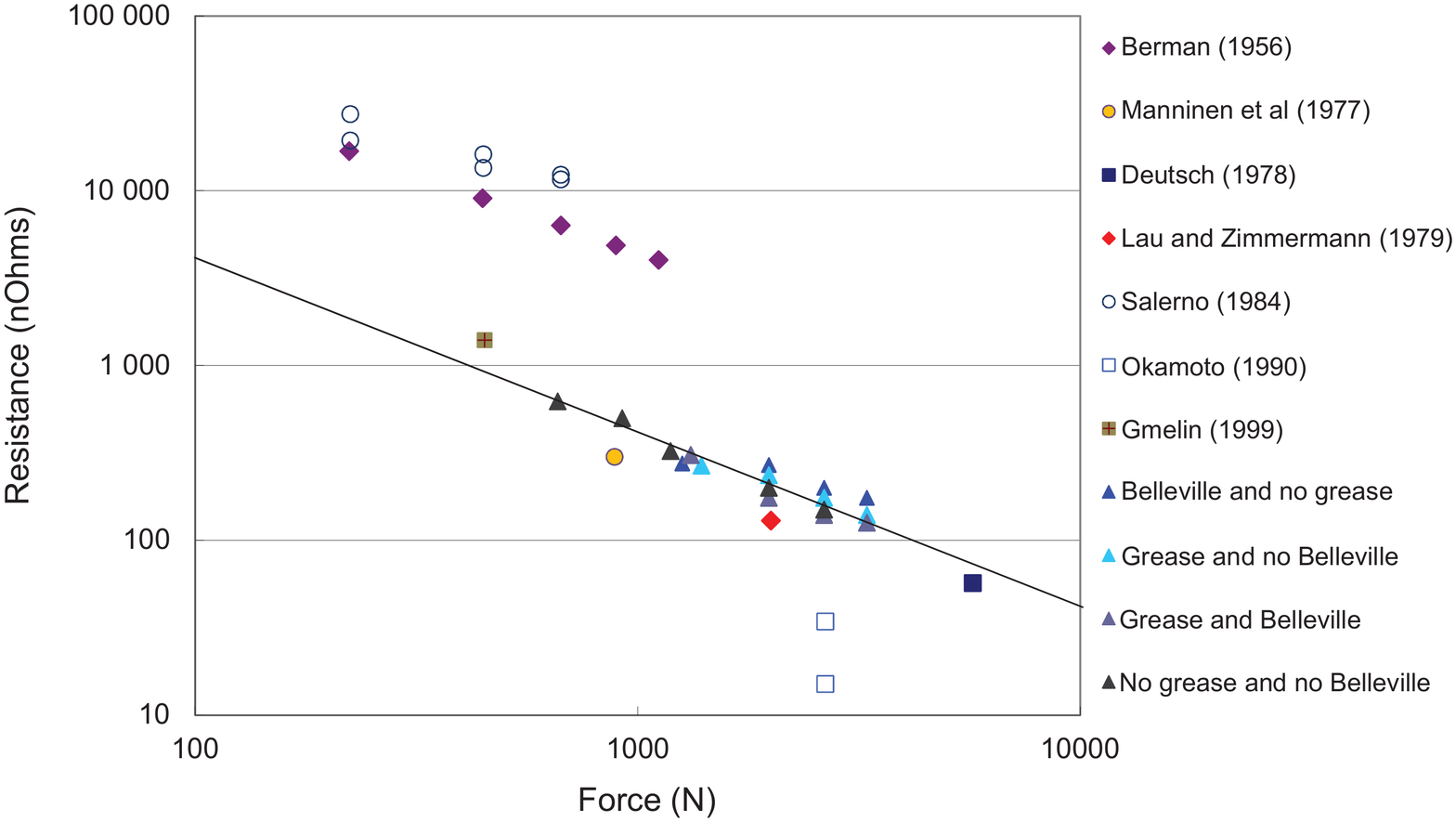}
\caption{Collection of experimental data for Cu-Cu contact resistances as function of applied force (from\cite{Blondelle2014}).}
\label{fig:Blondelle}       
\end{figure}

Numerous articles provide data on contact resistances between metals  (see for instance \cite{Blondelle2014,Schmitt2015,Didschuns2004,Salerno1997,Gmelin1999,Dhuley2019}). We limit ourselves to the value of electrical resistance at low temperatures and do not consider their –~sometimes non-linear~– temperature dependence, which is addressed in \cite{Dhuley2019}. We may identify 2 major trends in these data. A general trend is well represented in Fig.~\ref{fig:Blondelle} taken from Blondelle et al. \cite{Blondelle2014} and dealing with bolted Cu-Cu joints at 4.2~K. First, the resistances follow a $1/F$ dependence, where F is the applied Force, in agreement with theoretical predictions \cite{Dhuley2019}. Furthermore, except for the measurements of Okamoto et al. (see below), values in the range of 100-400~n$\rm \Omega$ are encountered for forces of 3000~N, close to the recommended force for standard M4 stainless steel bolts \cite{Blondelle2014}.

This is consistent with the ''conservative'' approximation proposed in \cite{Schmitt2015} for the thermal conductance of Cu-Cu bolted joints based on several earlier publications: $G = 0.0624\times T-0.00023$~[WK$^{-1}$]\footnote{It seems surprising that, in this article from Schmitt et al., the conductance would not depend on the applied force. We suppose that this approximation is valid for their conditions, i.e. 3~kN.}. This translates into a contact resistance Rc of $\approx$400~n$\rm \Omega$ using the Wiedemann-Franz law, which is comparable to the figure given above. Similar values are also found in \cite{Salerno1997} where Salerno and Kittel report thermal conductances for Cu assemblies (not only Cu-Cu) ranging from 10$^{-3}$ to $\approx$10$^{-1}$ WK$^{-1}$ at 4.2~K for 670~N applied force. With a linear dependence of the conductance with the applied force, then the best values at 3~kN would be close to 200~n$\rm \Omega$, again comparable to the figure given above.\\
Yet, a few authors report significantly lower values with different preparation techniques (we provide a detailed table in \nameref{sec:8}):

\begin{itemize}
\item Deutsch \cite{Deutsch1979} achieved 77~n$\rm \Omega$ by simply pressing copper samples one against the other.
\item Mueller et al. \cite{Mueller1978} and Shigematsu et al. \cite{Shigematsu1997} for Al-Cu contacts used rather complex procedures making use of a Zincate solution followed by cyanide-based gold plating. They achieved 5~n$\rm \Omega$ in the best conditions.
\item Okamoto et al. \cite{Okamoto1990} used non cyanide gold plating solutions.Their best results -- 4~n$\rm \Omega$ -- were obtained adding an indium filler.
\item Willekers et al. \cite{Willekers1989} used impact welding and achieved 32~n$\rm \Omega$.
\end{itemize}
Even if these measurements may appear exceptional compared to the large number of contact resistances above 100~n$\rm \Omega$, they suggest that a R$_c$ of 20~n$\rm \Omega$ for our Al/Cu contacts should be feasible.

Bad contact resistances with Al on one or 2 sides are often explained by the presence of a few nm Al$_2$O$_3$ hard film on the surface of aluminium. Such a hard oxide layer is difficult to get rid of and is detrimental for thermal contacts at very low temperatures. In the worst cases, the conductance of the contact would depart from the T dependence expected for metallic contacts \cite{Dhuley2019}. Several ways are proposed to remove or disrupt this oxide layer:

\begin{itemize}
\item Wanner \cite{Wanner1981} reports Al to Al contacts with an equivalent electrical resistance of 70~n$\rm \Omega$ for rough samples whereas polished samples give about 3 times this value. He suggests that the roughness combined with strong compression forces makes it possible to break the Al$_2$O$_3$ layer. The R$_c$ = 70~n$\rm \Omega$ was obtained for a large torque of 20~Nm onto a M8 screw. This R$_c$ is too high for our application and applying such a torque is out of range for our future system (we will use M4 screws at best).
\item We have performed tests using magnetic pulse welding inspired by developments in another field \cite{Raoelison2014} but the results were not satisfactory.
\item Impact welding, as successfully achieved by Willekers et al.\cite{Willekers1989} could be a promising approach, however it is unclear what effect this process might have on the highly annealed copper and aluminium required for our application.
\item It is often proposed to chemically remove the oxide layer and replace it by a coating (usually gold). The results are sometimes clearly unsatisfactory when the process is not appropriate \cite{Wanner1981} but some very good results have been obtained as in \cite{Mueller1978,Shigematsu1997}. In the chemical oxide removal, a zincate solution replaces  Al$_2$O$_3$ with Zn or Cu and the Al sample can be exposed to air. It is then gold plated in a separate solution.
\item An alternative solution proposed by Shigematsu et al. \cite{Shigematsu1997} drew our attention. It consists in using equipment dedicated to micro-electronics in order to sputter Au onto Al immediately after etching while remaining under vacuum. Despite their pessimistic conclusion on this approach, we have decided to investigate that process since we had already succeeded in removing the oxide layers on Al for other applications \cite{Goupy2018}. We have also decided to test Cu-Al connections with indium as a filler. 
\end{itemize}

For clarity, we first focus on our successful procedures and results, i.e. our last runs. Our earlier trial-and-error tests that may provide some useful information are reported in subsequent sections.

\section{Experimental setup}
\label{sec:3}
\begin{figure}[h!]
\centering
\includegraphics[width=0.99\linewidth]{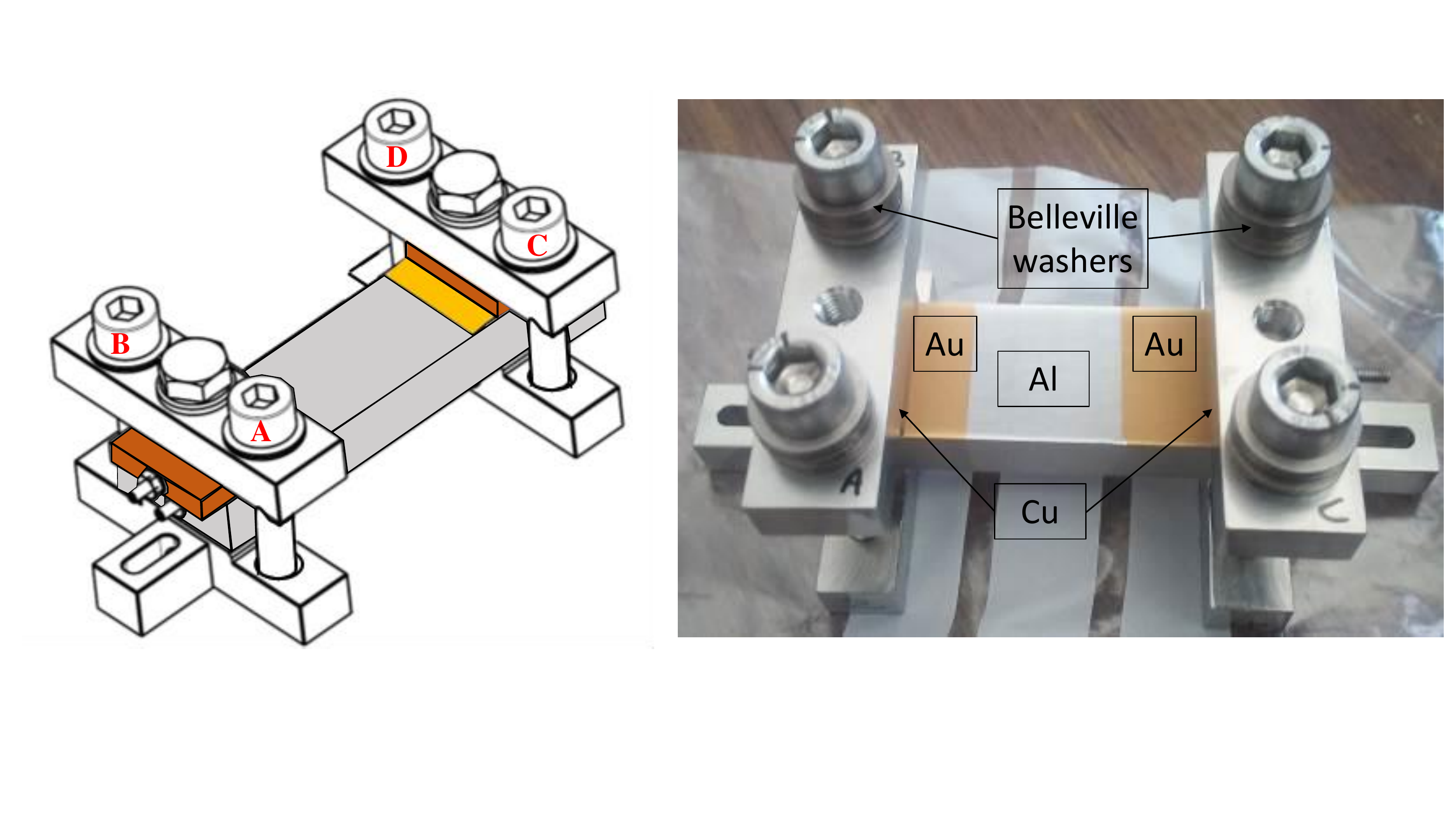}
\caption{Setup for resistivity measurements. Left: schematic drawing (Al in grey, Cu in orange, Au in yellow and SS in white). Right: assembly with an Al sample gold plated at the 2 ends. Note that Cu is not visible on the photo. Overall dimensions: $\approx 100~mm \times 50~mm \times 50~mm$.}
\label{fig:ResistSetup}
\end{figure}

\begin{figure}[h!]
\floatbox[{\capbeside\thisfloatsetup{capbesideposition={left,center},capbesidewidth=4cm}}]{figure}[\FBwidth]
{\caption{Sectional view of one Cu-Al assembly with the 4 + 4 Belleville washer configuration used for our measurements. The height ''h'' is used to determine the applied force. Kapton tape is inserted between the lower SS bar and the Al block in order to get a negligible current flow through the SS screw – SS lower bar – Al block path.}\label{fig:BellevilleSetup}}
{\includegraphics[width=9cm]{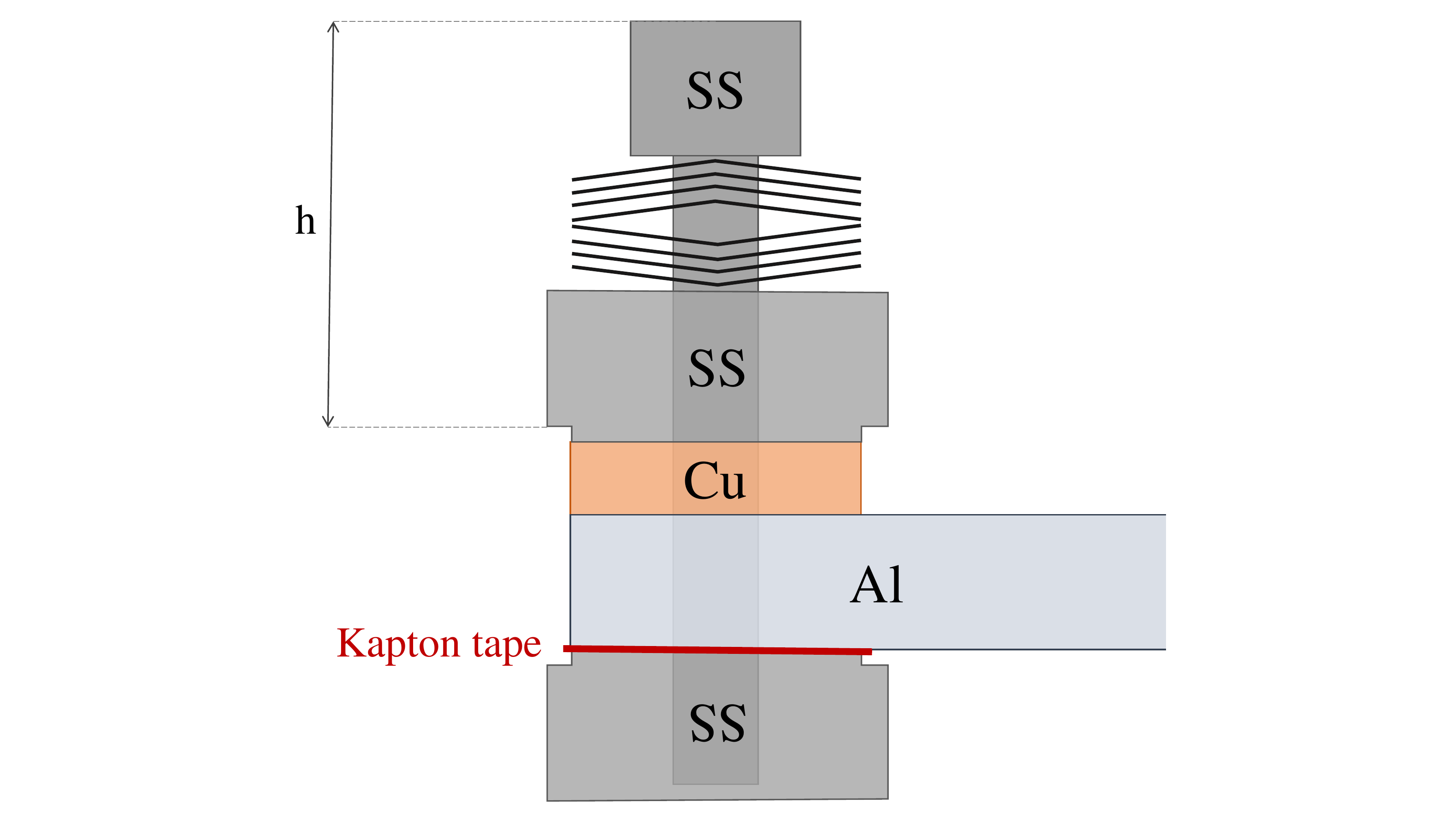}}
\end{figure}

The experimental setup for the samples to be tested is depicted in Fig. \ref{fig:ResistSetup} and \ref{fig:BellevilleSetup}. Small rectangular Cu pieces ($20~mm \times 12~mm \times 2.5~mm$) are pressed onto the 2 ends of a bigger Al block ($600~mm \times 20~mm \times 9.7~mm$) via 2 SS bars and 2 $\times$ 2 M6 SS screws + Belleville washers ($\varnothing$$_i$ = 6~mm,  $\varnothing$$_e$ = 12.5~mm, t = 0.7~mm, max. deflection 0.4~mm). Smaller screws are screwed into the different materials for current supply or voltage measurement.

In order to establish a reproducible and well controlled tightening procedure, we made some preliminary tests. First, we pressed representative Cu pieces against an Al block at different torques. For each test, we would put a pressure measurement film (Fuji Prescale MS) between the two materials. Although the film does not give very quantitative results, it helped us fix 5~Nm as a minimum torque and underlined the need for alternative tightening of the screws in order to get a significant and homogeneous contact surface. It also confirmed that our polishing procedure (see below) would give a good planarity.

Then, we tested and characterised several Belleville washer configurations. Belleville washers are needed to ensure that the applied force remains constant while the setup is being cooled down – about 10~$\rm \mu$m differential thermal contraction was expected for our device. The characterisation of the Belleville washers was judged necessary firstly because we could find very few technical data on compression characteristics of standard washers. We applied pressure on the screw head with a hydraulic press and measured the deflection of the washer assembly. The maximum measured stroke was $\approx$0.9~mm, close to the expected 0.8~mm. We could verify that the deflection increases linearly with the applied force. For our setup, the slope would be 0.23~mm.kN$^{-1}$ up to 3.5~kN. With this “calibration”, we could evaluate the force on our setup by measuring the deflection (''h'' in Fig. \ref{fig:BellevilleSetup}) instead of the torque applied on the M6 screws. This is far more reliable since, as detailed in \nameref{sec:8}, one cannot reliably convert a torque into a force. This latter observation was confirmed as we used different levels of cleaning for our M6 screws: similar applied torques would not compress the washers equivalently. This is explained by the variation of the SS-SS friction coefficient as a function of lubrication.

After assembly, the overall setup is mounted in a LHe cryostat.

\section{Samples preparation}
\label{sec:4}
We have used the following bulk materials which were not heat treated for our tests:

\begin{itemize}
\item Al plate ($9.7~mm \times 130~mm \times 370~mm$) - purity 4N6 \cite{Laurand}. On a test sample, we measured a RRR of $\approx$350 for the raw material.
\item Standard CuC2 ($\backsimeq$OFHC) block. On a test sample, we measured a RRR of $\approx$130 for the raw material.
\end{itemize}
2 Al (Al1, Al2) and 4 Cu (Cu1 to Cu4) samples were machined to the desired shape. They were then hand-polished using sequentially abrasive papers P400, P800, P2500, P4000, and finally 1~$\rm \mu$m diamond paste. A verification on a profilometer gives a roughness of $\approx$20~nm on the surface with some scratches in the 100 -- 500~nm range. There are more and deeper scratches for Al, for which a homogeneous polishing is difficult to achieve since particles are easily detached from the bulk during polishing.
After polishing, a first “soft” cleaning sequence has been applied: 3~minutes ultrasonic bath in RBS Neutral T -- 6\% concentration \cite{rbs} -- in hot water for Al and 3~minutes ultrasonic bath in RBS T 305 -- 5\% concentration \cite{rbs} -- in hot water for Cu, followed by 3~minutes ultrasonic rinsing bath in a beaker of hot water and 3~minutes ultrasonic bath in a beaker of ethanol, and finally drying with a soft cloth. The samples were then stored for days or weeks.\\
Immediately before Au evaporation or insertion of In, the samples underwent a stronger cleaning sequence:

\begin{itemize}
\item For Al and Cu: Remover (MicroChem 1112A \cite{microchem}) $\approx$15 mins + Rinse with de-ionized water.
\item For Al and Cu: Acetone 10~mins in ultrasonic bath + isopropyl alcohol 2~mins ultrasonic bath.
\item For Al only: a standard ''Aluminium Etch'' process as follows: 75\% H$_3$PO$_4$ + 2-5\% HNO$_3$ + 20\% de-ionized water $\approx$12 mins + rinse in de-ionized water until resistance of water is $>$12 M$\rm \Omega$. Note that the etching rate is expected to be around 20~nm/min resulting in the removal of up to $\approx$240~nm of Al$_2$2O$_3$. After this step, the surface was slightly less mirror-like and surface pitting was visible in some places.
\end{itemize}
Setup with gold plating -- sample reference Al1 + Cu3 (AB side) + Cu4 (CD side):

\begin{itemize}
\item About 40~minutes after cleaning, introduction of the 2 Cu pieces and Al block into an electron beam evaporator (Plassys model MEB550).
\item Argon plasma etch with the following parameters: I = 52~mA, V = 600~V, 3~minutes with the samples under rotation (5~rpm)\footnote{Very rough estimates based on measured etching rates of 7.4~nm/min for SiO$_2$ on Si and on \cite{IonMillRates} suggest an etching rate of $\approx$1.5~nm/min for Al$_2$O$_3$.} 
\item Electron beam evaporation of 200~nm of Au with the following parameters: 0.5~nm/s deposit (V = 9.85~kV, I = 308~mA, 5~rpm rotation, vacuum level $\approx$2 $\times$ 10$^{-5}$~Pa).
\item Immediately after removal from the evaporator, the two gold surfaces were placed in contact. Compressive force was applied by alternately tightening the bolts. 
\item Application of an effort of 11~kN with a hydraulic press (performed 2 days after sample assembly).
\item Re-tighten bolts to obtain a compression force of about 3 kN per bolt (5 Nm measured torque).
\end{itemize}
Setup with indium -- sample reference Al2 + Cu1 (AB side) + Cu2 (CD side):

\begin{itemize}
\item A 99.999\% purity piece of indium \cite{advent} approximately $1~mm \times 1mm \times 15mm$ was placed at the centre of the contact surface between the Al and Cu, which were then pressed together. A small sphere of indium was squeezed rather than a thin foil to limit the initial surface to volume ratio of the indium and consequently the quantity of oxide present on its surface.
\item Assembly of Al + Cu parts in the clamp and tightening of the bolts progressively to 8.5~Nm. The applied force was insufficient to fully squash the indium.
\item Application of an effort of 11~kN with a hydraulic press (performed 2 days after sample assembly).
\item Re-tighten bolts to obtain compression of the Belleville washers of $\approx$0.7~mm (3~kN per bolt) -- typical tightening torque 3.5 - 4.5~Nm
\end{itemize}

\section{Results}
\label{sec:5}
\begin{figure}[h!]
\centering
\includegraphics[width=0.95\linewidth]{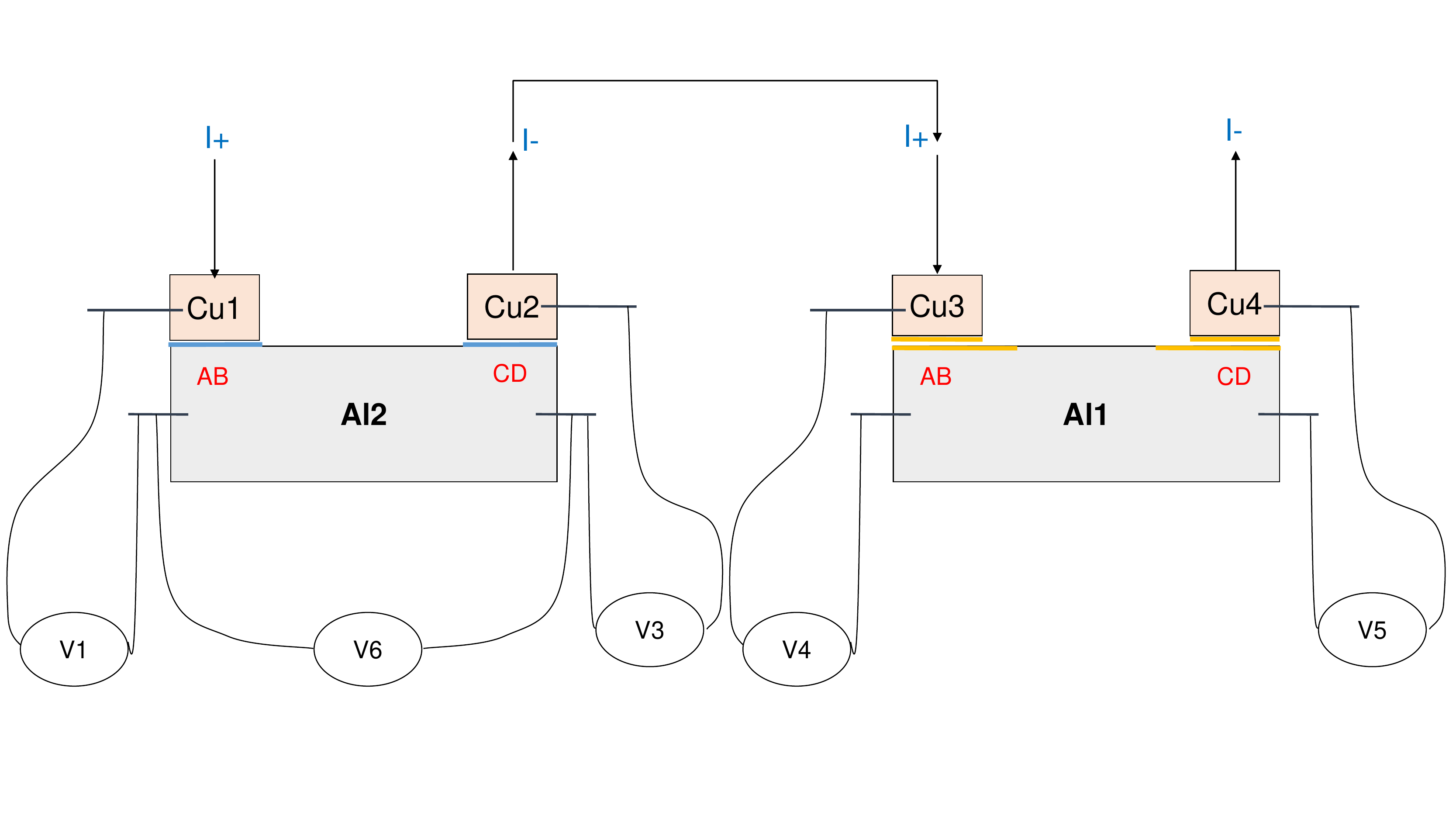}
\caption{Current and voltage pickups for the two samples (mounted in series). Left: with In filler (blue). Right: with Au deposit on Cu and Al (yellow).}
\label{fig:MeasSetup}
\end{figure}

Measurements are of 4 wire type. The two samples are mounted in series as presented in Fig. \ref{fig:MeasSetup}. The voltage is measured accross each of the contact resistances (V1, V3, V4 and V5). Actually, for each of these measurements, there is a contribution from bulk Cu and Al. From their measured RRRs, we evaluate their total contribution to be about 3~n$\rm \Omega$ for each of the two samples. Measurement V6 is used as a verification of the overall setup since the characteristics of the Al block are well known.

We used a program developed under LabView$^{\rm TM}$ to analyse the resistance of our samples. Each measurement sequence is achieved in several steps with current reversal between steps so as to compensate for thermo-electric effects. Each step lasts about 2~minutes and we programmed between 7 and 21 steps. The current (Kikusui Model PBZ40-10 current source) is chosen between 1 and 10~A. Copper current leads of low resistance run inside the cryostat and the current is injected into the samples at the level of the stainless steel support central screw (see Fig. \ref{fig:ResistSetup}). We always tested several current values in order to maximise the signal while checking for overheating effects. For each current step, voltages are measured sequentially using a Keithley model 2000 multimeter equipped with a multiplexer. For each channel, the voltage is averaged over about 20 measurements during each current step. The corresponding resistance is averaged over the current steps, ignoring the first and second steps, during which the system is stabilizing.

We always made a first set of measurements at 300~K so as to check all the connections, verify the overall measurement chain, compare with the calculated value for bulk Al and get first results on the contact resistances. The samples are then slowly cooled down to 4.2K during several hours using first liquid nitrogen and then liquid helium.

Results obtained with these 2 samples ($\Leftrightarrow$ 4 contacts) are summarized in Table \ref{tab:Synthesis}. Prior to the second measurement sequence at 4.2~K, the bolts A \& B on each sample (Cu1 \& Cu3) were loosened by approximately 0.4~mm corresponding to a final force of $\approx$1.3~kN for each bolt. The measurements were repeated 4 months later. Finally, the bolts A \& B on each sample were loosened to 0.2~mm compression of Belleville washers corresponding to a final force of $\approx$0.87~kN for each bolt. For this latest cool down, we also removed Belleville washers on bolts C \& D on each setup (Cu2 \& Cu4) then tightened back the bolts to 0.75~Nm, which corresponds to a force between 0.7 to 1~kN (friction coefficients between 0.18 and 0.12) for each bolt. We chose such low forces and also to remove Belleville washers in order to mimic the most simple cryogenic setup one can think of.

\begin{table}[h!]
\caption{Synthesis of resistance measurements.}
\centering
\begin{adjustbox}{max width=\textwidth}
	\begin{tabular}{c|c|c|c|c|c|c}
	\hline
	\textbf{Contact Ref.} &\textbf{ R (300~K)} & \textbf{R (4.2~K)} &\textbf{ RRR} & \textbf{R (4.2~K)} & \textbf{R (4.2~K)} & \textbf{R (4.2~K)}\\
	 & \textbf{ n$\rm \Omega$} & \textbf{n$\rm \Omega$} &  & \textbf{n$\rm \Omega$} & \textbf{n$\rm \Omega$} & \textbf{n$\rm \Omega$} \\
	\hline
	 & \textbf{28/11/19} & \textbf{29/11/19} &  & \textbf{14/01/20} & \textbf{20/05/20} & \textbf{29/05/20} \\
	Cu3-Au-Al1 & 274$\pm$73 & 3$\pm$6 & 91 & 4$\pm$3$^{\star}$ & 3.6$\pm$4 & 3.1$\pm$3$^{\dagger}$ \\
	Cu4-Au-Al1 & 297$\pm$62 & 2$\pm$7 & 149 & 4$\pm$4 & 3.2$\pm$3 & 4.7$\pm$3$^{\bullet}$ \\
	Cu1-In-Al2 & 715$\pm$63 & 12$\pm$8 & 60 & 14$\pm$3$^{\star}$ & 24$\pm$6 & 23.4$\pm$5$^{\dagger}$ \\
	Cu2-In-Al2 & 672$\pm$67 & 8$\pm$7 & 84 & 10$\pm$3 & 13$\pm$6 & 16$\pm$5$^{\bullet}$ \\
	\hline
	\end{tabular}
\end{adjustbox}
\begin{flushleft}
\footnotesize{$^{\star}$ Bolts loosened by about 0.4~mm} \\
\footnotesize{$^{\dagger}$ Bolts loosened to 0.2~mm compression of Belleville washers } \\
\footnotesize{$^{\bullet}$ Belleville washers removed and bolt tightened back to 0.75~Nm}
\end{flushleft}
\label{tab:Synthesis}
\end{table}

Whatever the type of junction, gold plated or with indium filler, the resistances at 4.2~K are close to or below 20~n$\rm \Omega$. The best results are obtained with gold plating. For the 2 gold plated samples, resistances cannot be distinguished from the expected contribution of bulk materials. Furthermore, no degradation with time is observed and there seems to be no dependence on the applied force, at least down to our minimum $\approx$1.5~kN total force. The 11~kN force applied with the press may have ''frozen'' the Al-Au-Cu interfaces so that it is not necessary to maintain a strong force on the contact area. 

The contact resistances are higher with indium. They do not seem to depend much on the applied force either, but may degrade with time. This latter effect is difficult to quantify due to measurement uncertainties.

\section{Complementary results and discussion}
\label{sec:6}
In this section, we give complenetary information deduced from previous measurements (with few details for clarity purpose):

\begin{itemize}
\item It has been suggested \cite{Salerno1997,Guillaume1990} that highly polished surfaces usually don’t give the lowest resistances. But, on the other hand, evaporated gold is expected to adhere better to a highly polished surface. We have tested roughly polished gold plated samples (typ. 1~$\rm \mu$m roughness for Al and Cu) but in all cases the resistance was above 100~n$\rm \Omega$, even after addition of indium as a filler. We cannot confirm that polishing in the 20-40~nm range is an optimum but rough samples do not give very low resistances.
\item For a sample undergoing a less agressive argon etching ($\approx$2.5 times less Al$_2$O$_3$ removed compared to previous sample) and experiencing a lower force ($\approx$6~kN via screwing instead 11~kN via hydraulic press), the contact resistance was above 230~n$\rm \Omega$ initially and degraded with time. We believe this to be caused by residual  Al$_2$O$_3$ which could then serve as starting point for further oxidation. Application of  8~kN via hydraulic press on a similar sample only reduced the resistance from  240~n$\rm \Omega$ to 140~n$\rm \Omega$.
\item We dismounted some of the samples for visual inspection. The force required to separate the polished and gold plated copper and aluminium pieces was not negligible (but still by hand). All the gold over the contact area had transferred to the copper block, leaving a bare aluminium surface. Note that in the case of the roughly polished sample, the parts separated spontaneously which suggests that a more intimate contact is obtained for highly polished surfaces.
\end{itemize}

\section{Conclusion}
\label{sec:7}

We have demonstrated that very low Al/Cu contact resistances at low temperature -- in the order of a few n$\rm \Omega$ -- can be achieved with rather simple methods: one with the use of an indium filler (and no gold plating) and one with gold plating in a standard evaporator for micro-electronics. The gold plating samples gave better results and seem to be more stable with time.\\
Furthermore, unlike indium, gold will not undergo any superconducting transition at low temperature, which might affect the thermal conductance of the joint.\\
The necessary elements for obtaining such low Al/Cu contact resistances seem to be:

\begin{itemize}
\item  A fine polishing of the Al and Cu surfaces.
\item The total removal of the Al$_2$O$_3$ oxide layers. These layers are partially removed by chemical etching. For gold plated samples, complete removal is achieved via an Ar etching followed by Au evaporation without air exposure in between. For samples using indium, we believe that friction between the aluminum and the indium when the latter is squashed between the smooth surfaces is sufficient to break the oxide layer.
\item The application of strong compression forces, but only temporarily.
\end{itemize}

We believe these contacts are not compatible with multiple disconnections. It is obvious for samples using indium since, after dismounting, we saw that the Al block was strongly imprinted. For the gold plated samples, the gold will probably remain stuck to the copper piece, thus allowing the aluminium surface to oxidize again.

It might be interesting to measure the thermal conductance of the Al/Cu contacts at low temperatures. First, this would provide further information on the applicability of the Wiedemann-Franz law for contacts at low temperature since there are conflicting results in the literature. On one side a discrepancy up to a factor of 2.5 has been reported \cite{Boughton1967} while discrepancies are much smaller for Gloos et al. \cite{Gloos1988} – see Table~1. This would also allow testing the effect of the indium filler far below its superconducting transition. For the purpose of the construction of the CNDR, similar tests with (Cu-Au)-(Au-Cu) contacts will be needed in order to keep the overall heat switch dismountable together with a resistance below the required value of 150~n$\rm \Omega$.

\begin{acknowledgements}
\begin{sloppypar}
We acknowledge support from the ERC StG grant UNIGLASS No.714692 and ERC CoG grant ULT-NEMS No. 647917. The research leading to these results has received funding from the European Union's Horizon 2020 Research and Innovation Programme, under Grant Agreement no 824109.\\
This work has been performed at the “Plateforme Technologique Amont” (PTA) of Grenoble.
\end{sloppypar}
\end{acknowledgements}

\newpage
\begin{center}
\section*{Appendix}
\label{sec:8}
\end{center}

\begin{table}[h!]
\caption{Some of the lowest published (sorted by publication date) Cu-Cu or Cu-Al contact resistances at low temperature.}
\centering
\begin{adjustbox}{max width=\textwidth}

	\begin{tabular}{c|c|p{2.5cm}|c|p{1.5cm}|p{1cm}|p{1cm}|p{2.5cm}}
	\hline
	 \textbf{Ref.} &  \textbf{Assy.} &  \textbf{Technology} &  \textbf{Variable} &  \textbf{Published Value} &  \textbf{Equiv. Res.} &  \textbf{Est. force$^{\star}$} &  \textbf{Comments} \\
	\hline
	\cite{Mueller1978} & Cu-Al & Al foils pressed against Cu foils + gold plating (zincate solution for Al) with $3 \times $M3 brass screws nearly to the tension at which they yield & Conductance & 21~$\rm \mu$WmK$^{-1}$ at 66~mK & 77~n$\rm \Omega$  & 2.7~kN & Measured at 66~mK \\
	\hline
	\cite{Deutsch1979} & Cu-Cu & 2 Cu blocks (10~mm diameter),  NC8-32 SS screw. 4~Nm. & Electrical R. & 57~n$\rm \Omega$ & 57~n$\rm \Omega$ & $\approx$3~kN & Measured at 4.2~K \\
	\hline
	\cite{Deutsch1979} & Cu-Cu & Same as above + indium filler & Electrical R. &5~n$\rm \Omega$  & 5~n$\rm \Omega$  & $\approx$3~kN & Same as above \\
	\hline
	\cite{Okamoto1990} & Cu-Cu & 2 gold plated copper disks (4~mm thick) pressed with SS bolt. No filler. 4~Nm &  Electrical R. & 10~n$\rm \Omega$  & 10~n$\rm \Omega$  & 5.3~kN & Measured at 4.2~K. Current decay method. \\
	\hline
	\cite{Okamoto1990} & Cu-Cu & Same as above + indium filler. 5~Nm &  Electrical R. & 4~n$\rm \Omega$  & 4~n$\rm \Omega$  & 6.6~kN & Same as above \\
	\hline
	\cite{Willekers1989} & Cu-Al & Cu (1~mm) - Al (0.5~mm) - Cu (1~mm) plates impact welded + etched Cu &  Thermal R. & $1.3/T$~KW$^{-1}$ & 32~n$\rm \Omega$  & & From 27~mK to 250~mK \\
	\hline
	\cite{Shigematsu1997} & Cu-Al & Al foil (0.1~mm) pressed onto Cu block. Use of a zincate solution (bondar dip) for Al & Electrical R. & 5~n$\rm \Omega$cm$^{-2}$ &  &  & Measured at 4.2~K \\
	\hline
	\end{tabular}
\end{adjustbox}

\begin{flushleft}
\footnotesize{$^{\star}$ Extracting a force from the applied torque seems not a trivial procedure. For example, Deutsch \cite{Deutsch1979} reports a torque of 4~Nm with a NC8-32 SS screw. The corresponding force may vary by more than a factor of 4 depending on the actual input parameters. First, there are various expressions for the calculation of the force and some might be erroneous. As an example, if one assumes that the parameter D from equation (8) provided in \cite{Blondelle2014} is equivalent to the parameter 0.5$\times$(d$_s$ + d$_h$) from equation (A1) in \cite{Dhuley2019}, then the pre-factor for this contribution differs by a factor of 2. We believe \cite{Dhuley2019} should preferably be considered. But, mainly, there is a large dispersion in the friction coefficients to be used in these equations. In \cite{Blondelle2014}, a friction coefficient of 0.12 is assumed while it is of 0.53 in \cite{Dhuley2019}. From our understanding, the first one is better adapted to SS-SS bolts while the second one corresponds to a dry SS-Cu connection. In reference [55] of Dhuley’s article, we also find that the friction coefficient for SS-304L to Cu would be of 0.23 in the worst case. Taking equation (A1) from Dhuley together with his parameters for UNC 8-32 screws (table 2 in \cite{Dhuley2019}), and a torque of 4~Nm one can calculate forces of 1.4~kN, 3~kN and 5.4~kN for µ = µ$_b$ = 0.12, 0.23 and 0.53 respectively. We believe µ = 0.23 is more realistic and we thus propose 3~kN equivalent force in the table.} \\
\end{flushleft}

\label{tab:Rc}
\end{table}


\end{document}